\pdfoutput=1
\documentclass[12pt]{iopart}
\usepackage[utf8]{inputenc}
\usepackage{iopams}
\usepackage{amsfonts}
\usepackage{graphicx}
\usepackage{hyperref}

\newcommand{\muG}{$\mu$Gal}

\begin{document}
\maketitle
\comment{No correction for the light propagation within the cube: comment on ``Relativistic theory of the falling retroreflector gravimeter''}
\author{V D Nagornyi}
\address{WSP USA Inc., One Penn Plaza, New York, NY 10019, USA}
\ead{vn2@member.ams.org}

\begin{abstract}
Although the equation of motion developed in the paper (Ashby 2018 \textit{Metrologia} \textbf{55} 1) depends on the parameters of the falling cube, such as depth and refraction index, the parameters are only associated with powers of time no greater than one, and so do not affect the acceleration.
The paper's correction due to the light propagation within the cube is therefore not supported by the equation of motion, and probably caused by omissions in data analysis.
The 'speed of light' component of the acceleration that follows from the equation, agrees with the results obtained by other authors.
\end{abstract}
%
%
\pagestyle{empty}
%
%
%
%
\section{The correction and some of its properties}
The paper \cite{ashby2018} presents a new substantial (about 7 \muG\footnote{1 \muG = $10^{-8}$ ms$^{-2}$}) correction for absolute gravimeters, caused by  propagation of light within the cube reflector. The correction is\footnote{formula (47), third line, of \cite{ashby2018}}
\begin{equation}
\label{formula47}
\delta g = - \gamma (D n - d),
\end{equation}
where $\gamma$ is the vertical gradient of gravity, $D$ is the 
cube depth from face to corner, $n$ is the refractive index of glass, $d=D/4$.
A quick look at the correction reveals a property not common  (on a laboratory scale) for the phenomena related to the light propagation: the complete dependence on the gravity gradient $\gamma$. In particular,
\begin{itemize}
    \item the correction disappears if there is no vertical gradient,
    \item the correction has different signs at sites located above and below the Earth's surface, based on the sign of the gradient.
\end{itemize}
In fact, the correction translates the gravity value to the point located above the cube's centre of mass at the distance $(Dn-d)$. As per the paper, the calculations by formula (\ref{formula47}) agree with the result of the trajectory fitting. We discuss here both ways the correction was obtained.
\section{No new acceleration perturbations in the equation of motion}
The trajectory fitting is based on the phase of the reflected signal at the point of recombination\footnote{formula (33) of \cite{ashby2018}}:
\begin{equation}
\label{formula33}
\eqalign{
\phi(T,0)=\frac{2( D n-d) \Omega}{c}
+\frac{2 \Omega Z_0}{c}
-\frac{2 (D n-d) \Omega V_0}{c^2}
-\frac{2 \Omega Z_0 V_0}{c^2}\\
+T(-\Omega +\frac{2\Omega V_0}{c}-\frac{2\Omega V_0^2}{c^2}+\frac{2\Omega g ( D n-d)}{c^2}+\frac{2 g \Omega Z_0}{c^2})\\
+T^2(-\frac{g \Omega}{c}+\frac{3 g \Omega V_0}{c^2})-\frac{g^2 \Omega T^3}{c^2}\\
+\gamma \Omega \bigg( T(-\frac{2 ( D n-d)Z_0}{c^2}-\frac{2 Z_0^2}{c^2})
+T^2(\frac{Z_0}{c}-\frac{(D n-d) V_0}{c^2}-\frac{4V_0Z_0}{c^2})\\
+T^3(\frac{V_0}{3c}
+\frac{g( D n-d)}{ 3c^2}-\frac{4 V_0^2}{3 c^2}
+\frac{7g Z_0}{3c^2})
+T^4(-\frac{g}{12 c}+\frac{5 g V_0}{4 c^2})-\frac{g^2 T^5}{4 c^2}  \bigg).
}
\end{equation}
By adding the phase $\Omega T$ of the reference beam and multiplying the resulting beat phase by 
$c/2\Omega$, we get the position of the cube at the moment $T$:
%
%
%
\begin{equation}
\label{formula33Z.1}
\eqalign{
Z(T) = (D n-d) + Z_0
-\frac{(D n-d) V_0}{c}
-\frac{Z_0 V_0}{c}\\
+T(V_0 -\frac{V_0^2}{c} + \frac{g (D n-d)}{c} + \frac{g Z_0}{c})\\
+T^2(-\frac{g}{2}+\frac{3g V_0}{2c})-\frac{g^2 T^3}{2 c}
+\gamma \bigg( 
\frac{Z_0 T^2}{2} +\frac{V_0 T^3}{6}
-\frac{g T^4}{24} \bigg).
}
\end{equation}
The terms with $\gamma$ represent the trajectory perturbation due to the vertical gravity gradient, the terms with $1/c$ represent the perturbation due to the finite speed of light. The factor $\gamma/c$ indicates a negligible cross-perturbation not shown here.
The second derivative of the coordinate yields the acceleration of the cube:
\begin{equation}
\label{formula33g}
\eqalign{
g(T) =  \underbrace{- g }_{\small \textrm{normal} \atop \textrm{ acceleration}}
+ \underbrace{\frac{3g}{c} (V_0 - g T)}_{\small \textrm{`speed of light'} \atop \textrm{perturbation}}
+\underbrace{\gamma \bigg( 
Z_0 + V_0 T
-\frac{g T^2}{2} \bigg)}_{\small \textrm{vertical gradient} \atop  \textrm{perturbation}}.
}
\end{equation}
The resulting acceleration consists of three familiar terms: the normal acceleration, the `speed of light' perturbation, and the vertical gradient perturbation. 
The equation reveals no perturbations related to the cube depth $D$.  Indeed, as the terms $(Dn-d)$ in (\ref{formula33}) are associated with powers of $T$ no greater than 1, they can only affect the estimates of the initial displacement $Z_0$ and velocity $V_0$. Therefore, the reported sensitivity of $g$ to the value of $D$ can only result from incorrect implementation of the fitting procedure, for example, from misplacing the components of (\ref{formula33Z.1}) in the regression matrix. 
\section{Non-identical polynomials cannot coincide}
To derive the formula for the correction, the following argument was used.
Let $Z_0, V_0, g$ be the estimates of the parameters in case $D=0$. Let some value of $D \ne 0$ change the estimates to $Z_0+\delta Z_0,  V_0 + \delta V_0, g + \delta g$. \textit{``All these quantities are constants independent of $T$, so the two phase functions at an arbitrary value of T must match.''}-- says the paper\footnote{between formulae (45) and (46)}. While the values are indeed independent of $T$, two polynomials with non-coinciding coefficients can only match at the number of points not exceeding the degree of the polynomials. In other words, two different constants never match, two different lines can match at most in 1 point, two quadratic parabolas -- in 2, cubic -- in 3, quartic (like the polynomial (\ref{formula33Z.1})) -- in 4. Therefore, the conclusion of two functions matching everywhere cannot serve as basis for deriving the correction.
\section{The 'speed of light' perturbation confirmed}
As seen in (\ref{formula33g}), the equation (\ref{formula33}) has all necessary time delays (terms with $1/c^2$) to account for the finite speed of light, even though the delays are not treated in \cite{ashby2018} explicitly. We agree with \cite{ashby2018} that using formula (\ref{formula33}) to get the acceleration requires no \textit{additional} `speed of light' correction.
The formulae (\ref{formula33g}) confirms that the 'speed of light' perturbation is proportional to the reflector's velocity with coefficient $3g/c$. Using the perturbation, one can find the correction \cite{nagornyi2011}, which needs to be applied to the result when no delays are retained in fitting.
\section{Conclusions}
While several attempts were previously made to describe absolute gravimeter in relativistic terms, the paper \cite{ashby2018} supercedes them in rigour and completeness of the analysis. The detailed tracking of the beam phase through the instrument, with full respect of relativistic laws and principles, culminates in the equation (\ref{formula33}). The problems only occur at fitting the  trajectory model to experimental data. The claimed correction for the light propagation within the cube disagrees with the equation of motion (\ref{formula33} -- \ref{formula33g}), which reveals no new acceleration components  On the other hand, the perturbation due to the finite speed of light that follows from the equation derived in \cite{ashby2018}, agrees with previous results of other authors. No change in the currently accepted ways of introducing the 'speed of light' correction is necessary. 
%
%
%
%
%
\section*{References}

\begin{thebibliography}{10}
%
\bibitem{ashby2018}
\href{http://iopscience.iop.org/article/10.1088/1681-7575/aa9ba1}
{
N Ashby 2018 Relativistic theory of the falling retroreflector gravimeter \textit{Metrologia} \textbf{55} 1
}
\bibitem{nagornyi2011}
\href{http://iopscience.iop.org/article/10.1088/0026-1394/48/3/004}
{
Nagornyi V D, Zanimonskiy Y M and Zanimonskiy Y Y 2011 Correction due to the finite speed of light in absolute gravimeters \textit{Metrologia} \textbf{48} 101–13
}
\end{thebibliography}
\end{document}